\documentclass[fleqn,usenatbib]{mnras}

\usepackage[T1]{fontenc}
\usepackage{ae,aecompl}

\usepackage{graphicx}	% Including figure files
\usepackage{amsmath}	% Advanced maths commands
\usepackage{amssymb}	% Extra maths symbols
\usepackage{bm}         % boldface symbols
\usepackage{color}      % use colours

\newcommand{\dd}{{\rm d}}
\newcommand{\pt}{{p_{\mathrm t}}}
\newcommand{\pg}{{p_{\mathrm g}}}
\def\Ldel{\delta}             % operator for the Lagrangian variation

\title[Surface effects]{On the surface physics affecting solar oscillation frequencies}

\author[G. Houdek et al.]{
G. Houdek,$^{1}$\thanks{E-mail: hg@phys.au.dk}
R. Trampedach,$^{2,1}$ %\thanks{E-mail: art@phys.au.dk}
M. J. Aarslev,$^{1}$
J. Christensen-Dalsgaard$^{1}$
\\
$^{1}$Stellar Astrophysics Centre, Department of Physics and Astronomy, Aarhus University, 8000 Aarhus C, Denmark\\
$^{2}$Space Science Institute, 4750 Walnut Street, Suite 205, Boulder, CO 80301, USA
}

\date{Accepted XXX. Received YYY; in original form ZZZ}

\pubyear{2016}

\begin{document}
\label{firstpage}
\pagerange{\pageref{firstpage}--\pageref{lastpage}}
\maketitle

\begin{abstract}
Adiabatic oscillation frequencies of stellar models, computed with 
the standard mixing-length formulation for convection, increasingly deviate with 
radial order from observations in solar-like stars. 
Standard solar models overestimate adiabatic frequencies by as much as $\sim$20$\mu$Hz.
In this letter, we address the physical processes of turbulent convection that are 
predominantly responsible for the frequency differences between standard models and 
observations, also called `surface effects'. 
We compare measured solar 
frequencies from the MDI instrument on the SOHO spacecraft with frequency calculations 
that include three-dimensional (3D) hydrodynamical simulation results 
in the equilibrium model,
nonadiabatic effects, and a consistent treatment of the turbulent pressure 
in both the equilibrium and stability computations.
With the consistent inclusion of the above physics in our model computation we are able
to reproduce the observed solar frequencies to $\lesssim$ 3$\mu$Hz without the need of
any additional ad-hoc functional corrections. 
\end{abstract}


%%%%%%%%%%%%%%%%%%%%%%%%%%%%%%%%%%%%%%%%%%%%%%%%%%

%%%%%%%%%%%%%%%%% BODY OF PAPER %%%%%%%%%%%%%%%%%%

\section{Introduction}

\begin{figure}
	\includegraphics[width=\columnwidth]{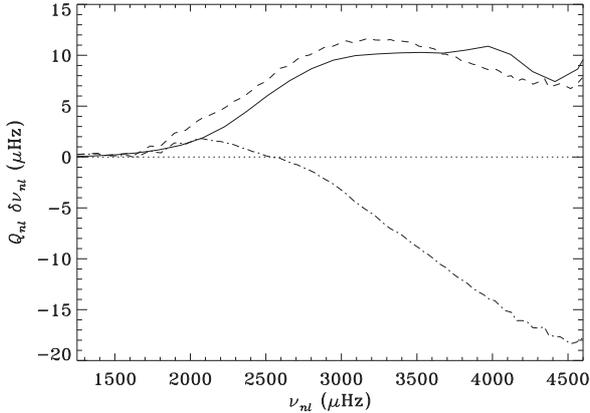}
    \caption{Inertia-scaled frequency differences between MDI measurements (Sun)
    of acoustic modes with degree $l=20$--$23$ and 
    model computations as functions of oscillation frequency.
    The scaling factor $Q_{nl}$ for a mode with radial order $n$,
    is obtained from taking ratios between the inertia of modes with $l=23$
    and 
    %the inertia of 
    radial modes, interpolated to the $l=23$ 
    frequencies \citep[e.g.,][]{AertsEtal10}. 
    % provided by the standard solar `Model S' \citep{JCDEtal96}. 
    The dot-dashed curve shows the differences for baseline model, 
    `Sun\,-\,{A' (cf. Section~\ref{sec:1Dbmodel})},
    reflecting the results for a standard solar model computation. 
    The dashed curve plots the residuals for the patched model which 
    includes turbulent pressure and convective back-warming in the 
    mean model, i.e. `Sun\,-\,{B} (cf. Section~\ref{sec:31Dmodel})'. 
    The solid curve illustrates the differences from
    the modal effects of nonadiabaticity and perturbation to the turbulent pressure, i.e.
    `{D\,-\,C}'\, (cf. Section~\ref{sec:1Dmodel}).}
    \label{fig:MDI-1D.a-1DPM.a-NL}
\end{figure}

{High-quality measurements of} stellar oscillation frequencies 
of {thousands} of solar-like stars 
are now available from the NASA space mission $Kepler$, and from the French satellite
mission CoRoT (Convection Rotation and planetary Transits).
In order to exploit these 
data for
probing stellar interiors, accurate modelling of stellar oscillations is required.
However, 
adiabatically computed  
frequencies are increasingly overestimated with 
increasing radial order (see dot-dashed curve in Fig.~\ref{fig:MDI-1D.a-1DPM.a-NL}). 
These effects have become known as `surface effects' 
\citep[e.g.,][]{Brown84, Gough84, Balmforth92b,
RosenthalEtal99, Houdek10, GrigahceneEtal12}.
Semi-empirical corrections to adiabatically
computed frequencies proposed 
by e.g., \citet{KjeldsenEtal08}, \citet{BallGizon14}, \citet{SonoiEtal15}, \citet{BallEtal16}
are purely descriptive and provide little physical insight. 
Here, we report on a self-consistent model computation, which reproduces the observed
solar frequencies to within $\sim3\mu$Hz, and, for the first time, without the need 
of any ad-hoc functional corrections. It represents a purely physical explanation for 
the `surface effects' by considering (a) a state-of-the-art 3D--1D patched %models for the
mean model, (b) nonadiabatic effects, (c) a consistent treatment of turbulent 
pressure in the mean and pulsation models, and (d) a depth-dependent modelling of the 
turbulent anisotropy in both the mean and oscillation calculations. 

Convection modifies pulsation properties of stars principally through three effects: 
\begin{itemize}
\item[(i)]effects through the turbulent pressure term
in the {hydrostatic equation} ({structural} effect), and its pulsational 
perturbation in the momentum equation (modal effect);

\item[(ii)]opacity variations brought about by large 
convective temperature fluctuations, affecting the mean stratification; this {structural} effect 
is also known as `convective back-warming' \citep[e.g.,][]{TrampedachEtal13};
 
\item[(iii)]nonadiabatic effects, additional to the pulsational perturbed radiative heat flux, 
through the perturbed convective heat flux (modal effects) in the 
thermal energy equation. % \citep[e.g.,][]{HoudekDupret15}.
\end{itemize}

We follow \citeauthor{RosenthalEtal99}'s~(\citeyear{RosenthalEtal99}) idea
of replacing the outer layers of a 1D solar envelope model by an averaged 
3D simulation, 
and adopt the most advanced and accurate 3D -- 1D matching procedure
available today 
\citep{TrampedachEtal14a, TrampedachEtal14b}
for estimating {structural} effects on adiabatic solar frequencies.
Furthermore, we use
a 1D nonlocal, time-dependent convection model for estimating the modal effects 
of nonadiabaticity and convection dynamics.

Additional modal effects can be associated with the advection of the oscillations by
spatially varying turbulent flows in the limit of temporal stationarity 
\citep{Brown84, ZhugzhdaStix94, BhattacharyaEtal15}. This `advection picture' is related 
to the `dynamical picture' of including temporally varying turbulent convection 
fluctuations in the limit of spatially horizontal homogeneity. 
Because these two pictures describe basically the same effect but in two 
different limits, i.e. they are complementary,
only one of them should be included. We do so by adopting the latter.

\vspace{-5pt}
\section{Model computations}
\label{sec:models}

We use stellar envelope models in which the total pressure $p=p_{\rm g}+p_{\rm t}$
satisfies the equation for hydrostatic support,
\begin{equation}
    \frac{\dd p}{\dd m}=-\frac{1}{4\pi r^2}\frac{Gm}{r^2}\,,
	\label{eq:hydstat}
\end{equation}
where $\pg$ is the gas pressure and $\pt$ the turbulent
pressure, $\pt:=\overline{\rho ww}$, with $w$ being the vertical component of the
convective velocity field $\bm{u} = (u,v,w)$, and an overbar indicates an ensemble average.
The other symbols are mass $m$, radius $r$, mass density $\rho$, and
gravitational constant $G$.

\vspace{-3pt}
\subsection{Adiabatic pulsations of mean models constructed with turbulent pressure}
\label{sec:adcalturb}
If turbulent pressure is included in the model's mean stratification,
particular care must be given to frequency calculations that neglect the 
pulsational perturbation to the turbulent pressure.
In an adiabatic treatment 
the relative gas pressure perturbation $\Ldel\pg/\pg$ is related to 
the relative density perturbation $\Ldel\rho/\rho$ by the linearized expression
\begin{equation}
   \frac{\Ldel\rho}{\rho}
   =\frac{1}{\gamma_1}\frac{\Ldel\pg}{\pg}
   =\frac{1}{\gamma_1}\frac{p}{\pg}\left(\frac{\Ldel p}{p}-\frac{\Ldel\pt}{p}\right)\,,
   \label{eq:perturbed-adiab-eos}
\end{equation}
where $\Ldel X(m)$ {are perturbations following the motion, and} 
$\gamma_1:=(\partial\ln\pg/\partial\ln\rho)_{s}$ is the first adiabatic exponent 
with $s$ being specific entropy. 
A standard adiabatic calculation typically neglects convection dynamics, i.e. the 
effect of the perturbation to the turbulent pressure, $\Ldel\pt/p$, leading to the approximate 
linearized expression for an adiabatic change

\begin{equation}
   \frac{\Ldel\rho}{\rho}
   \simeq\frac{1}{\gamma_1}\frac{p}{\pg}\frac{\Ldel p}{p}\,.
   \label{eq:perturbed-adiab-eos2}
\end{equation}

\begin{figure}
    \includegraphics[width=\columnwidth]{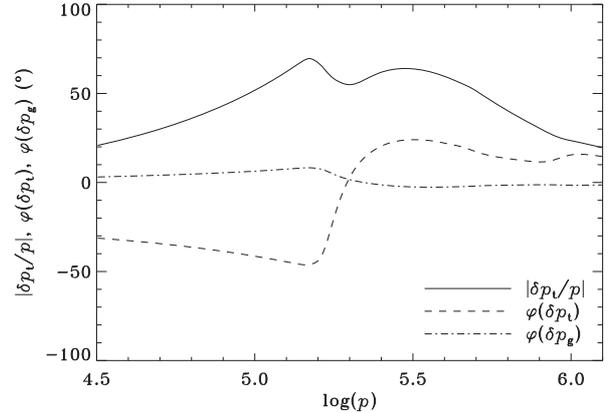}
    \caption{Norm of the relative turbulent pressure eigenfunction $|\Ldel\pt/p|$ (solid curve)
    and phases of $\Ldel\pt$ (dashed curve) and
    gas pressure perturbations $\Ldel\pg$ (dot-dashed curve){;}  
    {the relative displacement eigenfunction is normalized to unity at the surface.}
    Results are shown for a radial
    mode with frequency $\nu\simeq 2947\,\mu$Hz, obtained with the 
    solar envelope model `{D}' of Section~\ref{sec:1Dmodel}.}
    \label{fig:dpt-phase}
\end{figure}

Neglecting $\Ldel\pt$ in the full expression~(\ref{eq:perturbed-adiab-eos}) for an adiabatic change
is partially justified from full nonadiabatic pulsation calculations, in which the turbulent
pressure and its pulsational perturbation, $\Ldel\pt$,
are consistently included.
Such a pulsation computation, in which the 
pulsational perturbation to the convective heat flux and turbulent pressure
%and in which $\Ldel\pt$ 
is obtained from a time-dependent convection 
model \citep[e.g.,][]{HoudekDupret15},
shows that $\Ldel\pt$ varies predominantly in quadrature with % the gas pressure 
perturbation $\Ldel\pg$. This is illustrated in Fig.~\ref{fig:dpt-phase}, for 
the 1D solar envelope model computed according to Section~\ref{sec:1Dmodel}, 
where the phases $\varphi(\Ldel\pt)$ (dashed curve) and $\varphi(\Ldel\pg)$ 
(dot-dashed curve) of turbulent and gas pressure perturbations are plotted as a 
function of the total pressure $p$ for a particular radial mode.
The solid curve is the norm 
$|\Ldel\pt/p|$ of the relative turbulent pressure eigenfunction. In layers where
$|\Ldel\pt/p|$ is largest, the difference between
$\varphi(\Ldel\pt)$ and $\varphi(\Ldel\pg)$ can be as large as
$\sim 60\degr$, indicating that the turbulent pressure perturbation contributes
predominantly to the imaginary part of the complex eigenfrequency, i.e. to the damping 
or driving
of the pulsation modes.

Equation~(\ref{eq:perturbed-adiab-eos2}) describes consistently, in 
view of the $\pt-$term in the hydrostatic equation~(\ref{eq:hydstat}), 
the approximation of neglecting $\Ldel\pt$ in the adiabatic frequency calculations.
Therefore, if turbulent pressure is included in the stellar equilibrium structure, the
only modification to the adiabatic oscillation equations is the inclusion of the 
factor $p/\pg$ in the expression for an adiabatic change~(\ref{eq:perturbed-adiab-eos2})
\citep[see also][]{RosenthalEtal99}.
{
Omitting this factor is inconsistent with neglecting $\Ldel\pt$ in the 
adiabatic frequency calculations. 
}

\begin{figure}
	\includegraphics[width=\columnwidth]{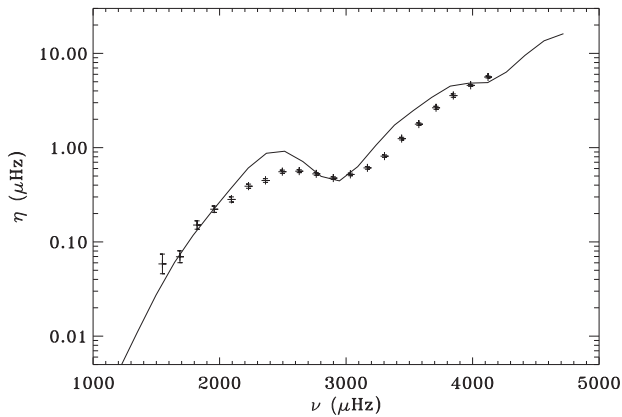}
    \caption{Radial damping rates in units of cyclic frequency of model `{D}' 
    (values are connected by solid lines) are
    compared with BiSON measurements of half the linewidths in the spectral peaks of the
    observed solar power spectrum (symbols with error bars; from \citealt{ChaplinEtal05}).}
    \label{fig:BiSON.HWHM-1D.eta}
\end{figure}

\subsection{Envelope models}
The convection effects on the mean model structure are investigated by comparing
envelope models, computed with either the standard mixing-length formulation 
or by adopting appropriately averaged 3D simulation results for the outer layers of 
the convective envelope,
with solar frequencies measured by the MDI\footnote{Michelson Doppler Imager} 
instrument \citep{ScherrerEtal95} 
on the SOHO\footnote{SOlar and Heliospheric Observatory} spacecraft.
The modal effects are estimated by comparing adiabatic and nonadiabatic frequencies from
1D envelope models constructed with a nonlocal, time-dependent formulation for the 
mean and pulsational perturbations to the 
convective heat flux and turbulent pressure.

The adopted models are  

\begin{description}
\item[{\bf A}:] adiabatically computed oscillations of a 1D baseline model constructed
with the standard mixing-length formulation \citep{BohmVitense58} for convection.

\item[{\bf B}:] adiabatically computed oscillations of a 
patched model that was constructed
by replacing the outer parts of the convection zone of the baseline model, BM, 
by averaged hydrodynamical simulation results. It therefore includes the 
turbulent pressure and the effect of 
convective back-warming in the mean model.
The turbulent pressure perturbation, $\Ldel\pt$, 
is omitted in the adiabatic oscillation calculations according to 
equation~(\ref{eq:perturbed-adiab-eos2}).

\item[{\bf C}:] adiabatically computed oscillations of a 
1D nonlocal mixing-length model including
turbulent pressure $\pt$ in the equation of hydrostatic support, but omitting 
convective back-warming and the effect of the
pulsational Lagrangian perturbations to the turbulent pressure %, $\Ldel\pt$, 
in the adiabatic oscillation calculations according to equation~(\ref{eq:perturbed-adiab-eos2}). 

\item[{\bf D}:] nonadiabatically 
computed oscillations of the same 1D nonlocal mean model 
used for {model `C'}, including the Lagrangian perturbations to turbulent pressure, 
$\Ldel\pt$, and to the convective heat flux.
\end{description}

\begin{figure}
	\includegraphics[width=\columnwidth]{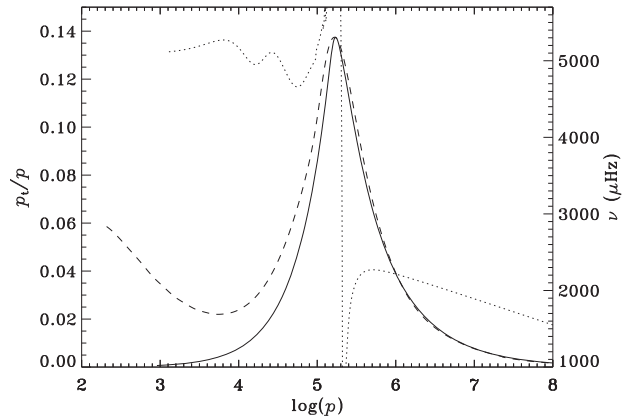}
    \caption{
    Comparison of the turbulent pressure over total pressure between
    the patched mean model `{B}' (dashed curve), for which the convection zone was
    modelled by averaged 3D simulation results, and the calibrated nonlocal mean model 
    `{D}' (solid curve), as functions of the logarithmic total pressure.
    The dotted curve is the acoustic cutoff frequency \citep[e.g.,][]{AertsEtal10} indicating
    the region of mode propagation ($\log p\gtrsim 5.3$).
    }
    \label{fig:max-pt}
\end{figure}

\subsubsection{The 3D convective atmosphere simulation}
\label{sec:3Dsim}
The 3D simulation, {described by \citet{TrampedachEtal13}}, evolves the conservation
equations of mass, momentum and energy on a regular grid, which is optimized
in the vertical direction to capture the photospheric transition. 
The equation of state (EOS) is
a custom calculation of \citeauthor{MihalasEtal88}'s~(\citeyear{MihalasEtal88}) 
EOS  for the employed 15 element mixture,
and the monochromatic opacities are described by \citet{TrampedachEtal14a}.
Radiative transfer is solved explicitly with the hydrodynamics, and 
line-blanketing (non-greyness) is accounted for by a binning of the 
monochromatic opacities, as developed by \citet{Nordlund82}. The top and bottom
boundaries are open and transmitting, minimizing their effect on the interior
of the simulation. The constant entropy assigned to the inflows at the bottom 
is adjusted to obtain the \hbox{solar effective temperature $T_{\rm eff}$.}

%\vspace{-5pt}
\subsubsection{Baseline model - {\rm `A'}}
\label{sec:1Dbmodel} 
The baseline model is a 1D solar envelope model integrated from
a Rosseland optical depth of $\tau=10^{-4}$ down to a depth of 5\%\,R$_\odot$,
and using \citeauthor{BohmVitense58}'s~(\citeyear{BohmVitense58}) mixing-length 
formulation for convection.
The model is computed with a code \citep{JCDFrandsen83} that is closely related
to \citeauthor{JCD08}'s~(\citeyear{JCD08}) stellar evolution code ASTEC. The
turbulent pressure is omitted in the hydrostatic equation~(\ref{eq:hydstat}). 
The 1D model adopts the same atomic physics
as the 3D atmosphere simulation described above in Section~\ref{sec:3Dsim}.
The 3D simulation also provides the temperature-optical-depth relation 
and the mixing length for the 1D baseline model 
\citep{TrampedachEtal14a, TrampedachEtal14b}. 
This is accomplished by matching the 1D baseline model to the 3D simulation at a 
common pressure sufficiently deep that the 3D convective fluctuations can be
considered linear and far enough from the bottom of the 3D spatial domain
that boundary effects are negligible.
The total mass and
luminosity are identical for the 1D baseline model and the 3D simulation, whereas
$T_{\rm eff}$ and surface gravity are diluted in the latter case by the convective
expansion of the 3D atmosphere, which also gives rise to the stratification
part of the `surface effects' in the patched model of Section~\ref{sec:31Dmodel}. 
The limited extent of the envelope models restricts our mode selection 
to those that have lower turning points well inside the lower boundary. 
Choosing modes with degree
$l=20$--$23$ fulfils this requirement, as well as ensures that the modes are
predominantly radial on the scale of the thin layer giving rise to the
surface effects.

%\vspace{-5pt}
\subsubsection{Patched model with $\pt$ - {\rm `B'}}
\label{sec:31Dmodel} 
Since the 1D baseline model is matched continuously to 
the 3D simulation (see Section~\ref{sec:1Dbmodel}),
the two solutions can be combined to a single, patched, model for the 
adiabatic oscillation calculations. This, however, demands one more step: the 
3D simulation is carried out in the plane-parallel approximation, and their constant
gravitational acceleration introduces significant glitches in some quantities.
We therefore apply a correction for sphericity, consistent 
with the \hbox{radius of the 1D model}.

%\vspace{-5pt}
\subsubsection{Nonlocal models with $\pt$ - {\rm `C \& D'}}
\label{sec:1Dmodel}
The 1D nonlocal model calculations with turbulent pressure are carried out 
essentially in the manner described by 
\citet[][see also \citealt{Balmforth92a}]{HoudekEtal99}. 
The convective heat flux
and turbulent pressure are obtained from a 
nonlocal generalization of the mixing-length formulation \citep{Gough77a, Gough77b}. 
In this generalization three more parameters, $a, b$ and $c$, are introduced which
control the spatial coherence of the ensemble of eddies contributing
to the total convective heat flux ($a$) and turbulent pressure ($c$), and the 
degree to which the turbulent fluxes are coupled to the local stratification ($b$).
The effects of varying these nonlocal parameters on the solar structure and oscillation
properties were discussed in detail by \citet{Balmforth92a}.

The nonlocal parameter $c$ is calibrated such as to have the
maximum value of the turbulent pressure, max($\pt$), in the 1D nonlocal model 
to agree with the 3D simulation result 
(see Fig.~\ref{fig:max-pt}). 
The depth-dependence of the anisotropy 
$\Phi:=\bm{u}\cdot\bm{u}/{w^2}$ of the convective velocity field 
is adapted from the 3D simulations using an analytical function 
with the maximum 3D value in the 
atmospheric layers and the minimum 3D value
in the deep interior of the simulations.
The {remaining} nonlocal parameters $a$ and $b$ 
{cannot be easily obtained from the 3D simulations and}
are therefore calibrated such as to have a good
agreement between calculated damping rates and measured solar linewidths
(see Fig.~\ref{fig:BiSON.HWHM-1D.eta}). 
The mixing length was calibrated to the helioseismically determined
convection-zone depth $d_{\rm cz}/{\text R}_\odot\simeq0.287$ \citep{JCD-DOG-MJT91}.
Both the envelope and pulsation calculations assume the generalized Eddington
approximation to radiative transfer \citep{UnnoSpiegel66}.
The abundances by mass of hydrogen and heavy elements are adopted from the
patched model `{B}', i.e. $X=0.736945$ and $Z=0.018055$.
The opacities are obtained from the OPAL tables
\citep{IglesiasRogers96}, supplemented at low temperature by tables 
from \citet{Kurucz91}.
The EOS includes a detailed treatment of the ionization
of C, N, and O, and a treatment of the first ionization of the next
seven most abundant elements \citep{JCD82}.
The integration of stellar-structure equations starts 
at an optical depth of $\tau=10^{-4}$ and ends at a radius fraction $r/$R$_\odot=0.2$.
The temperature gradient in the plane-parallel atmosphere is corrected by 
using a radially varying Eddington factor fitted to Model C of 
\citet{VernazzaEtal81}.

The linear nonadiabatic pulsation calculations are carried out using
the same nonlocal convection formulation with the assumption that all eddies in the cascade
respond to the pulsation in phase with the dominant large eddies.
A simple thermal outer boundary condition is adopted at the temperature minimum 
where for the mechanical boundary condition the solutions are 
matched smoothly onto those of a plane-parallel isothermal atmosphere 
\citep[e.g.,][]{BalmforthEtal01}.
At the base of the model envelope the 
conditions of adiabaticity and vanishing of the displacement eigenfunction 
are imposed. Only radial p modes are considered.

\begin{figure}
	\includegraphics[width=\columnwidth]{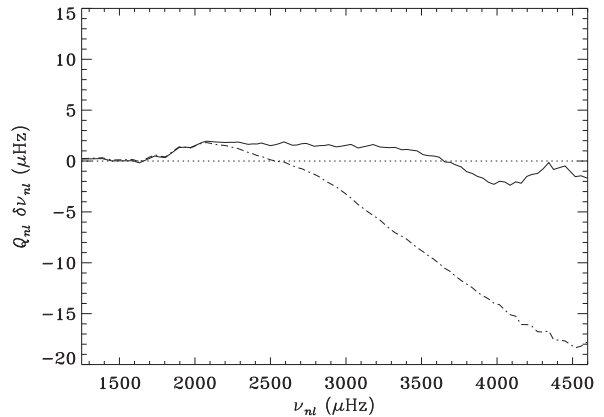}
    \caption{Inertia-scaled frequency difference between MDI data (Sun) and model 
    calculations. The solid curve includes the
    combined frequency corrections arising from {structural} effects (`{B}')
    and modal effects (`{D}'). 
    The dot-dashed curve is the result for our baseline model `{A}', reflecting
    the result for a `standard' solar model computation.}
    \label{fig:MDI-1D[all].na}
\end{figure}

%\vspace{-5pt}
\section{Results and discussion}

The adiabatic frequency corrections (Section~\ref{sec:adcalturb}) arising 
from modifications to the stratification of the mean model are obtained from an 
appropriately averaged 3D simulation for the outer
convection layers (Section~\ref{sec:31Dmodel}).

The frequency corrections associated with modal effects arising from
nonadiabaticity, including both the perturbations to the radiation 
and convective heat flux, and from convection-dynamical effects of the perturbation
to the turbulent pressure, are estimated
from a 1D nonlocal, time-dependent convection model including turbulent pressure 
(Section~\ref{sec:1Dmodel}).

%\vspace{-5pt}
\subsection{Adiabatic frequency corrections from modifications to the mean structure}
\label{sec:mean_structure}
Frequency differences between MDI data (Sun) and our baseline model, `Sun - {A}',  
are depicted in Fig.~\ref{fig:MDI-1D.a-1DPM.a-NL} by the dot-dashed curve, illustrating the
well-known `surface effects' for a standard solar model with a frequency 
residual up to $\sim20\,\mu$Hz.
The effect on the adiabatic frequencies by adopting an averaged 3D simulation 
for the outer convection layers 
is illustrated by the dashed curve in Fig.~\ref{fig:MDI-1D.a-1DPM.a-NL}. It 
shows the frequency difference between MDI data and the patched model, 
`Sun - {B}'. The patched model underestimates the frequencies
by as much as $\sim10\,\mu$Hz. The change from overestimating the frequencies 
with the baseline model, `{A}' (dot-dashed curve), to underestimating the 
frequencies with the patched model, `{B}' (dashed curve),
is mainly due to effects of turbulent pressure $\pt$ in the equation of hydrostatic 
support~(\ref{eq:hydstat}) and from opacity changes (convective back-warming) of the 
relatively large convective temperature fluctuations in the superadiabatic boundary layers.   
\subsection{Modal effects from nonadiabaticity and convection dynamics}
\label{sec:model_effects}
Additional to the {structural} changes
we also consider the modal effects 
of nonadiabaticity and pulsational perturbation to turbulent pressure $\Ldel\pt$.
We do this by using the 1D solar envelope model of Section~\ref{sec:1Dmodel}, which includes
turbulent pressure, and which is calibrated such as to have the same max($\pt$) as the
3D solar simulation (see Fig.~\ref{fig:max-pt}).
To assess the modal effects we compute for this nonlocal envelope model 
nonadiabatic and adiabatic frequencies. 
The frequency differences between these two model computations, 
i.e. `{D\,-\,C}', is plotted in Fig.~\ref{fig:MDI-1D.a-1DPM.a-NL}
with a solid curve, and illustrates the modal effects of nonadiabaticity and turbulent 
pressure perturbations $\Ldel\pt$.
These modal effects (solid curve in Fig.~\ref{fig:MDI-1D.a-1DPM.a-NL}) produce 
frequency residuals that are similar 
in magnitude to the frequency residuals between the Sun and the patched 
model, `Sun\,-\,{B}' (dashed curve). This suggests that the underestimation 
of the adiabatic frequencies due to changes in the mean model, `{B}', is nearly 
compensated by the modal effects. The remaining overall frequency difference between the
Sun and models that include both {structural} and modal effects, i.e. the difference between 
the dashed and solid curves in Fig.~\ref{fig:MDI-1D.a-1DPM.a-NL}, is illustrated 
in Fig.~\ref{fig:MDI-1D[all].na} by the solid
curve, showing a maximum frequency difference of $\sim3\,\mu$Hz. 
Also depicted, for comparison, is the dot-dashed curve from Fig.~\ref{fig:MDI-1D.a-1DPM.a-NL},
which shows the frequency difference for the baseline model {`A'}, representing
the result for a standard solar model calculation.
 
We conclude that, 
if both {structural} and modal effects due to convection and nonadiabaticity are
considered together, it is possible to reproduce the measured solar frequencies satisfactorily
(solid curve in Fig.~\ref{fig:MDI-1D[all].na})
without the need of any ad-hoc correction functions.
Moreover, the calibrated set of convection parameters in the 
1D nonlocal model calculations 
reproduces the {turbulent-pressure profile 
of the 3D simulation in the relevant wave-propagating layers} (Fig.~\ref{fig:max-pt}), 
the correct depth of the convection zone, and
solar linewidths {over the whole measured frequency range} (Fig.~\ref{fig:BiSON.HWHM-1D.eta}).
Although we have not used the same equilibrium model for estimating the {structural}
(`{B}') and the modal effects (`{D}'), we believe that this remaining 
inconsistency is minute on the estimated modal effects, because of the satisfactory
reproduction of the $\pt$ profile in the nonlocal equilibrium model `{D}' 
(see Fig.~\ref{fig:max-pt}). 
However, we do plan to address this in a future paper.  

\section*{Acknowledgements}

We thank Douglas Gough for many inspiring discussions.
RT acknowledges funding from NASA grant NNX15AB24G.
Funding for the Stellar Astrophysics Centre is provided
by The Danish National Research Foundation (Grant DNRF106).

%%%%%%%%%%%%%%%%%%%%%%%%%%%%%%%%%%%%%%%%%%%%%%%%%%



%\bibliographystyle{mnras}
%\bibliography{GHoudekR2clean_arxiv} % your bibtex file xxxx.bib

\begin{thebibliography}{}
\makeatletter
\relax
\def\mn@urlcharsother{\let\do\@makeother \do\$\do\&\do\#\do\^\do\_\do\%\do\~}
\def\mn@doi{\begingroup\mn@urlcharsother \@ifnextchar [ {\mn@doi@}
  {\mn@doi@[]}}
\def\mn@doi@[#1]#2{\def\@tempa{#1}\ifx\@tempa\@empty \href
  {http://dx.doi.org/#2} {doi:#2}\else \href {http://dx.doi.org/#2} {#1}\fi
  \endgroup}
\def\mn@eprint#1#2{\mn@eprint@#1:#2::\@nil}
\def\mn@eprint@arXiv#1{\href {http://arxiv.org/abs/#1} {{\tt arXiv:#1}}}
\def\mn@eprint@dblp#1{\href {http://dblp.uni-trier.de/rec/bibtex/#1.xml}
  {dblp:#1}}
\def\mn@eprint@#1:#2:#3:#4\@nil{\def\@tempa {#1}\def\@tempb {#2}\def\@tempc
  {#3}\ifx \@tempc \@empty \let \@tempc \@tempb \let \@tempb \@tempa \fi \ifx
  \@tempb \@empty \def\@tempb {arXiv}\fi \@ifundefined
  {mn@eprint@\@tempb}{\@tempb:\@tempc}{\expandafter \expandafter \csname
  mn@eprint@\@tempb\endcsname \expandafter{\@tempc}}}

\bibitem[\protect\citeauthoryear{{Aerts}, {Christensen-Dalsgaard}  \&
  {Kurtz}}{{Aerts} et~al.}{2010}]{AertsEtal10}
{Aerts} C.,  {Christensen-Dalsgaard} J.,   {Kurtz} D.,  2010, Asteroseismology.
Springer

\bibitem[\protect\citeauthoryear{{Ball} \& {Gizon}}{{Ball} \&
  {Gizon}}{2014}]{BallGizon14}
{Ball} W.~H.,  {Gizon} L.,  2014, \aap\ %{10.1051/0004-6361/201424325},
  %\href {http://adsabs.harvard.edu/abs/2014A%26A...568A.123B} 
{568, A123}

\bibitem[\protect\citeauthoryear{{Ball}, {Beeck}, {Cameron}  \& {Gizon}}{{Ball}
  et~al.}{2016}]{BallEtal16}
{Ball} W.~H.,  {Beeck} B.,  {Cameron} R.~H.,   {Gizon} L.,  2016, 
  \aap\ %{10.1051/0004-6361/201628300}, %\href {http://adsabs.harvard.edu/abs/2016A%26A...592A.159B} 
  {592, A159}

\bibitem[\protect\citeauthoryear{{Balmforth}}{{Balmforth}}{1992a}]{Balmforth92a}
{Balmforth} N.~J.,  1992a, \mnras\ %\href {http://adsabs.harvard.edu/abs/1992MNRAS.255..603B} 
{255, 603}

\bibitem[\protect\citeauthoryear{{Balmforth}}{{Balmforth}}{1992b}]{Balmforth92b}
{Balmforth} N.~J.,  1992b, \mnras\ %\href {http://adsabs.harvard.edu/abs/1992MNRAS.255..632B} 
{255, 632}

\bibitem[\protect\citeauthoryear{{Balmforth}, {Cunha}, {Dolez}, {Gough}  \&
  {Vauclair}}{{Balmforth} et~al.}{2001}]{BalmforthEtal01}
{Balmforth} N.~J.,  {Cunha} M.~S.,  {Dolez} N.,  {Gough} D.~O.,   {Vauclair}
  S.,  2001, \mnras\ %{10.1046/j.1365-8711.2001.04182.x}, %\href {http://adsabs.harvard.edu/abs/2001MNRAS.323..362B} 
  {323, 362}

\bibitem[\protect\citeauthoryear{{Bhattacharya}, {Hanasoge}  \&
  {Antia}}{{Bhattacharya} et~al.}{2015}]{BhattacharyaEtal15}
{Bhattacharya} J.,  {Hanasoge} S.,   {Antia} H.~M.,  2015, \apj\ 
  %{10.1088/0004-637X/806/2/246}, %\href {http://adsabs.harvard.edu/abs/2015ApJ...806..246B} 
  {806, 246}

\bibitem[\protect\citeauthoryear{{B{\"o}hm-Vitense}}{{B{\"o}hm-Vitense}}{1958}]{BohmVitense58}
{B{\"o}hm-Vitense} E.,  1958, Zeitschrift f{\"u}r Astrophysik\ %\href {http://adsabs.harvard.edu/abs/1958ZA.....46..108B} 
  {46, 108}

\bibitem[\protect\citeauthoryear{{Brown}}{{Brown}}{1984}]{Brown84}
{Brown} T.~M.,  1984, Science\ %{10.1126/science.226.4675.687}, %\href {http://adsabs.harvard.edu/abs/1984Sci...226..687B} 
{226, 687}

\bibitem[\protect\citeauthoryear{{Chaplin}, {Houdek}, {Elsworth}, {Gough},
  {Isaak}  \& {New}}{{Chaplin} et~al.}{2005}]{ChaplinEtal05}
{Chaplin} W.~J.,  {Houdek} G.,  {Elsworth} Y.,  {Gough} D.~O.,  {Isaak} G.~R.,
   {New} R.,  2005, \mnras\ %{10.1111/j.1365-2966.2005.09041.x}, %\href  {http://adsabs.harvard.edu/abs/2005MNRAS.360..859C} 
   {360, 859}

\bibitem[\protect\citeauthoryear{{Christensen-Dalsgaard}}{{Christensen-Dalsgaard}}{1982}]{JCD82}
{Christensen-Dalsgaard} J.,  1982, \mnras\ %{10.1093/mnras/199.3.735},
  %\href {http://adsabs.harvard.edu/abs/1982MNRAS.199..735C} 
{199, 735}

\bibitem[\protect\citeauthoryear{{Christensen-Dalsgaard}}{{Christensen-Dalsgaard}}{2008}]{JCD08}
{Christensen-Dalsgaard} J.,  2008, \apss\ %{10.1007/s10509-007-9675-5},
  %\href {http://adsabs.harvard.edu/abs/2008Ap%26SS.316...13C} 
	  {316, 13}

\bibitem[\protect\citeauthoryear{{Christensen-Dalsgaard} \&
  {Frandsen}}{{Christensen-Dalsgaard} \& {Frandsen}}{1983}]{JCDFrandsen83}
{Christensen-Dalsgaard} J.,  {Frandsen} S.,  1983, \solphys\ %
  %{10.1007/BF00145588}, %\href {http://adsabs.harvard.edu/abs/1983SoPh...82..469C} 
  {82, 469}

\bibitem[\protect\citeauthoryear{{Christensen-Dalsgaard}, {Gough}  \&
  {Thompson}}{{Christensen-Dalsgaard} et~al.}{1991}]{JCD-DOG-MJT91}
{Christensen-Dalsgaard} J.,  {Gough} D.~O.,   {Thompson} M.~J.,  1991,
  \apj\ %{10.1086/170441}, 
  %\href {http://adsabs.harvard.edu/abs/1991ApJ...378..413C} 
  {378, 413}

\bibitem[\protect\citeauthoryear{{Gough}}{{Gough}}{1977a}]{Gough77b}
{Gough} D.~O.,  1977a, in {Spiegel} E.~A.,  {Zahn} J.-P.,  eds,  Lecture Notes
  in Physics Vol. 71, Problems of Stellar Convection. Springer, Heidelberg, 
  pp~15--56%, {10.1007/3-540-08532-7\_31}

\bibitem[\protect\citeauthoryear{{Gough}}{{Gough}}{1977b}]{Gough77a}
{Gough} D.~O.,  1977b, \apj\ %{10.1086/155244}, %\href {http://adsabs.harvard.edu/abs/1977ApJ...214..196G} 
{214, 196}

\bibitem[\protect\citeauthoryear{{Gough}}{{Gough}}{1984}]{Gough84}
{Gough} D.~O., 1984, Advances in Space Research\ %
  %{10.1016/0273-1177(84)90370-3}, %\href {http://adsabs.harvard.edu/abs/1984AdSpR...4...85G} 
  {4, 85}

\bibitem[\protect\citeauthoryear{{Grigahc{\`e}ne}, {Dupret}, {Sousa},
  {Monteiro}, {Garrido}, {Scuflaire}  \& {Gabriel}}{{Grigahc{\`e}ne}
  et~al.}{2012}]{GrigahceneEtal12}
{Grigahc{\`e}ne} A.,  {Dupret} M.-A.,  {Sousa} S.~G.,  {Monteiro}
  M.~J.~P.~F.~G.,  {Garrido} R.,  {Scuflaire} R.,   {Gabriel} M.,  2012,
  \mnras\ %{10.1111/j.1745-3933.2012.01233.x}, %\href {http://adsabs.harvard.edu/abs/2012MNRAS.422L..43G} 
  {422, L43}

\bibitem[\protect\citeauthoryear{{Houdek}}{{Houdek}}{2010}]{Houdek10}
{Houdek} G.,  2010, Astronomische Nachrichten
  %{10.1002/asna.201011445}, %\href {http://adsabs.harvard.edu/abs/2010AN....331..998H} 
  {331, 998}

\bibitem[\protect\citeauthoryear{{Houdek} \& {Dupret}}{{Houdek} \&
  {Dupret}}{2015}]{HoudekDupret15}
{Houdek} G.,  {Dupret} M.-A.,  2015, Living Rev. in Solar Phys.
  %{10.1007/lrsp-2015-8}, %\href {http://adsabs.harvard.edu/abs/2015LRSP...12....8H} 
  {12}

\bibitem[\protect\citeauthoryear{{Houdek}, {Balmforth}, {Christensen-Dalsgaard}
   \& {Gough}}{{Houdek} et~al.}{1999}]{HoudekEtal99}
{Houdek} G.,  {Balmforth} N.~J.,  {Christensen-Dalsgaard} J.,   {Gough} D.~O.,
  1999, \aap\ %\href {http://adsabs.harvard.edu/abs/1999A%26A...351..582H} 
  {351, 582}

\bibitem[\protect\citeauthoryear{{Iglesias} \& {Rogers}}{{Iglesias} \&
  {Rogers}}{1996}]{IglesiasRogers96}
{Iglesias} C.~A.,  {Rogers} F.~J.,  1996, \apj\ %{10.1086/177381},
  %\href {http://adsabs.harvard.edu/abs/1996ApJ...464..943I} 
{464, 943}

\bibitem[\protect\citeauthoryear{{Kjeldsen}, {Bedding}  \&
  {Christensen-Dalsgaard}}{{Kjeldsen} et~al.}{2008}]{KjeldsenEtal08}
{Kjeldsen} H.,  {Bedding} T.~R.,   {Christensen-Dalsgaard} J.,  2008, 
  \apjl\ %{10.1086/591667}, %\href {http://adsabs.harvard.edu/abs/2008ApJ...683L.175K} 
  {683, L175}

\bibitem[\protect\citeauthoryear{{Kurucz}}{{Kurucz}}{1991}]{Kurucz91}
{Kurucz} R.~L.,  1991, in {Crivellari} L.,  {Hubeny} I.,   {Hummer} D.~G.,
  eds,  NATO Series C Vol. 341, NATO (ASI). p.~441

\bibitem[\protect\citeauthoryear{{Mihalas}, {Dappen}  \& {Hummer}}{{Mihalas}
  et~al.}{1988}]{MihalasEtal88}
{Mihalas} D.,  {Dappen} W.,   {Hummer} D.~G.,  1988, \apj\ %
  %{10.1086/166601}, %\href {http://adsabs.harvard.edu/abs/1988ApJ...331..815M}
  {331, 815}

\bibitem[\protect\citeauthoryear{{Nordlund}}{{Nordlund}}{1982}]{Nordlund82}
{Nordlund} {\AA}., 1982, \aap\ %\href {http://adsabs.harvard.edu/abs/1982A%26A...107....1N} 
	  {107, 1}

\bibitem[\protect\citeauthoryear{{Rosenthal}, {Christensen-Dalsgaard},
  {Nordlund}, {Stein}  \& {Trampedach}}{{Rosenthal}
  et~al.}{1999}]{RosenthalEtal99}
{Rosenthal} C.~S.,  {Christensen-Dalsgaard} J.,  {Nordlund} {\AA}.,  {Stein}
  R.~F.,   {Trampedach} R.,  1999, \aap\ %\href {http://adsabs.harvard.edu/abs/1999A%26A...351..689R} 
	  {351, 689}

\bibitem[\protect\citeauthoryear{{Scherrer} et~al.,}{{Scherrer}
  et~al.}{1995}]{ScherrerEtal95}
{Scherrer} P.~H.,  et~al., 1995, \solphys\ %{10.1007/BF00733429}, %\href {http://adsabs.harvard.edu/abs/1995SoPh..162..129S} 
  {162, 129}

\bibitem[\protect\citeauthoryear{{Sonoi}, {Samadi}, {Belkacem}, {Ludwig},
  {Caffau}  \& {Mosser}}{{Sonoi} et~al.}{2015}]{SonoiEtal15}
{Sonoi} T.,  {Samadi} R.,  {Belkacem} K.,  {Ludwig} H.-G.,  {Caffau} E.,
  {Mosser} B.,  2015, \aap\ %{10.1051/0004-6361/201526838}, %\href {http://adsabs.harvard.edu/abs/2015A%26A...583A.112S} 
	  {583, A112}

\bibitem[\protect\citeauthoryear{{Trampedach}, {Asplund}, {Collet}, {Nordlund}
  \& {Stein}}{{Trampedach} et~al.}{2013}]{TrampedachEtal13}
{Trampedach} R.,  {Asplund} M.,  {Collet} R.,  {Nordlund} {\AA}.,   {Stein}
  R.~F.,  2013, \apj\ %{10.1088/0004-637X/769/1/18}, %\href {http://adsabs.harvard.edu/abs/2013ApJ...769...18T} 
  {769, 18}

\bibitem[\protect\citeauthoryear{{Trampedach}, {Stein},
  {Christensen-Dalsgaard}, {Nordlund}  \& {Asplund}}{{Trampedach}
  et~al.}{2014a}]{TrampedachEtal14a}
{Trampedach} R.,  {Stein} R.~F.,  {Christensen-Dalsgaard} J.,  {Nordlund}
  {\AA}.,   {Asplund} M.,  2014a, \mnras\ %{10.1093/mnras/stu889},
  %\href {http://adsabs.harvard.edu/abs/2014MNRAS.442..805T} 
  {442, 805}

\bibitem[\protect\citeauthoryear{{Trampedach}, {Stein},
  {Christensen-Dalsgaard}, {Nordlund}  \& {Asplund}}{{Trampedach}
  et~al.}{2014b}]{TrampedachEtal14b}
{Trampedach} R.,  {Stein} R.~F.,  {Christensen-Dalsgaard} J.,  {Nordlund}
  {\AA}.,   {Asplund} M.,  2014b, \mnras\ %{10.1093/mnras/stu2084},
  %\href {http://adsabs.harvard.edu/abs/2014MNRAS.445.4366T} 
  {445, 4366}

\bibitem[\protect\citeauthoryear{{Unno} \& {Spiegel}}{{Unno} \&
  {Spiegel}}{1966}]{UnnoSpiegel66}
{Unno} W.,  {Spiegel} E.~A.,  1966, \pasj\ %\href {http://adsabs.harvard.edu/abs/1966PASJ...18...85U} 
  {18, 85}

\bibitem[\protect\citeauthoryear{{Vernazza}, {Avrett}  \& {Loeser}}{{Vernazza}
  et~al.}{1981}]{VernazzaEtal81}
{Vernazza} J.~E.,  {Avrett} E.~H.,   {Loeser} R.,  1981, \apjs\ %
  %{10.1086/190731}, %\href {http://adsabs.harvard.edu/abs/1981ApJS...45..635V}
  {45, 635}

\bibitem[\protect\citeauthoryear{{Zhugzhda} \& {Stix}}{{Zhugzhda} \&
  {Stix}}{1994}]{ZhugzhdaStix94}
{Zhugzhda} Y.~D.,  {Stix} M.,  1994, \aap\ % 
%\href {http://adsabs.harvard.edu/abs/1994A%26A...291..310Z} 
{291, 310}

\makeatother
\end{thebibliography}

%\begin{thebibliography}{99}
%\end{thebibliography}

%%%%%%%%%%%%%%%%%%%%%%%%%%%%%%%%%%%%%%%%%%%%%%%%%%

%%%%%%%%%%%%%%%%% APPENDICES %%%%%%%%%%%%%%%%%%%%%

%\appendix



%%%%%%%%%%%%%%%%%%%%%%%%%%%%%%%%%%%%%%%%%%%%%%%%%%

% Don't change these lines
\bsp	% typesetting comment
\label{lastpage}
\end{document}